\documentstyle[mprocl]{article}

\input{psfig}
\def\Journal#1#2#3#4{{#1} {\bf #2}, #3 (#4)}

\newcommand{\beq}{\begin{equation}}
\newcommand{\eeq}{\end{equation}}
\newcommand{\bd}{\begin{displaymath}}
\newcommand{\ed}{\end{displaymath}}
\def\bea{\begin{eqnarray}}
\def\eea{\end{eqnarray}}

\def\ba{\beq\new\begin{array}{c}}
\def\ea{\end{array}\eeq}

\def\inbar{\,\vrule height1.5ex width.4pt depth0pt}
\def\IC{\relax\hbox{$\inbar\kern-.3em{\rm C}$}}
\def\IR{\relax{\rm I\kern-.18em R}}
\def\IN{\relax{\rm I\kern-.18em N}}

\def\Tr{{\rm Tr}}

\def\e{~{\rm e}}

\begin{document}

\title{The Mixed Non-Abelian Coulomb Gas in Two Dimensions
\footnote{Talk presented by G.W.S. at the Conference on Recent 
Developments in Non-Perturbative Quantum Field Theory (AIJIC 97),
May 26-30 1997 at Seoul, South Korea.}}

\author{ C.R. Gattringer, L.D. Paniak and G.W. Semenoff }

\address{Department of Physics and Astronomy,\\
University of British Columbia\\6224 Agricultural Road\\
Vancouver, British Columbia, Canada V6T 1Z1}   
 

\maketitle\abstracts{
The statistical mechanics of a mixed gas of adjoint and 
fundamental representation charges interacting via
1+1-dimensional $U(N)$ gauge fields is investigated.
In the limit of large $N$ we show that there is a first order
deconfining phase transition for low densities of
fundamental charges.  As the density of fundamental charges
becomes comparable to the adjoint charge density the 
phase transition becomes a third order one.}

\section{Introduction}
The classical Coulomb gas is an important model in statistical
mechanics.  It is exactly solvable in one dimension. In two dimensions
it exhibits the Berezinsky-Kosterlitz-Thouless phase transition which
is the prototype of all phase transitions in two-dimensional systems
which have $U(1)$ symmetry.  In this paper, we shall discuss a
generalization of the classical Coulomb gas to a system of quarks
which interact with each other through non-Abelian electric fields.
This model is known to be exactly solvable in some special cases, for
$SU(2)$ gauge group and fundamental representation quarks in one
dimension~\cite{nambu} and for $SU(N)$ gauge group in the large-$N$
limit with adjoint representation quarks in one dimension~\cite{stz,sz}.  
It can also be formulated on the lattice and solved
with adjoint quarks in the large $N$ limit in higher dimensions~\cite{sz,Zar} 
where it has a substantially more complicated
structure, although, even there, a solution of some special cases of
the model are relevant to the deconfinement transition of three and
four-dimensional Yang-Mills theory~\cite{dadda}.

In the present paper, we shall concentrate on solving a more general
version of the one-dimensional model than has previously been
considered and elaborating on the properties of the solution.  The 
one-dimensional case has the advantage of being directly related to a
continuum field theory, the heavy quark limit of 1+1-dimensional
quantum chromodynamics (QCD).  Part of the motivation for this work is
to study the possibility of a confinement-deconfinement phase
transition at high temperature or density in a field theory which has
some of the features of QCD, i.e.  with similar gauge symmetries and
where interactions are mediated by non-Abelian gauge fields.  QCD
exhibits confinement at low temperature and density with elementary
excitations being color neutral particles - mesons and baryons.  On
the other hand, at high temperature or high density it is very
plausible that the dynamical degrees of freedom would be quarks and
gluons - which would form a quark-gluon plasma, rather than the mesons
and baryons of low temperature nuclear physics.  At some intermediate
temperature or density there should be a crossover between these two
regimes.  There are few explicitly solvable models where this behavior
can be studied directly.  Previous to the model of~\cite{stz,sz},
known explicit examples of phase transitions in Yang-Mills theory were
not associated with confinement, but were either lattice artifacts~\cite{gw}  
unrelated to the continuum gauge theory or were
associated with topological degrees of freedom in Yang-Mills theory on
the sphere~\cite{dougkaz} and cylinder~\cite{dadda,matgro} and
demark a finite range of coupling constants within which the gauge
theory resembles a string theory.

In finite temperature Yang-Mills theory (or QCD with only adjoint
quarks), confinement is thought to be governed by the realization of a
global symmetry which is related to the center of the gauge group and
implemented by certain topologically non-trivial gauge transformations
that appear only at finite temperature~\cite{pol,sus,sy,sy2}.  The
Polyakov loop operator is an order parameter for spontaneous breaking
of this center symmetry and yields a mathematical way of
distinguishing the confining and deconfined phases.  When fundamental
representation quarks are present, the center symmetry is broken
explicitly and the Polyakov loop operator is no longer a good order
parameter for confinement.  Whether, in this case, a mathematical
distinction of confined and deconfined phases exists, and indeed
whether there is a distinct phase transition at all, is an open
question.

Here, we will consider a toy model which resembles two-dimensional QCD
with heavy adjoint and fundamental representation quarks.  It could
also be thought of as the heavy quark limit of dimensionally reduced
higher dimensional QCD where the adjoint particles are the gluons of
the compactified dimension which get a mass 
(similar to a Debye mass) from the dimensional reduction, 
and the fundamental representation
particles are the quarks.  We will solve this model explicitly in the
large-$N$ limit.  The model with only adjoint quarks was solved in
refs.~\cite{stz,sz} and it was found that the explicit solution has a
first order phase transition between confining and deconfining phases.
These phases could be distinguished by the expectation value of the
Polyakov loop operator, which provided an order parameter for
confinement in that case.  In this paper, we shall add fundamental
representation quarks. Then, as in QCD, the center symmetry is
explicitly broken and the Polyakov loop is always non-vanishing.  We
nevertheless find that the first order phase transition persists when
the density of fundamental quarks is sufficiently small.  When the
density of fundamental quarks is increased until it is comparable to
the density of adjoint quarks, the phase transition becomes a second
order one.  When the fundamental quark density is increased further,
the phase transition is third order.
 
Another motivation of the present paper, as well as Ref.~\cite{stz,sz}
is to study a suggestion by Dalley and Klebanov~\cite{dk} and
Kutasov~\cite{kutasov} that 
1+1-dimensional adjoint QCD would be the simplest gauge theory model which
exhibits some of the stringy features of a confining gauge theory.  It
is a long-standing conjecture that the confining phase of a gauge
theory can be described by a string theory~\cite{polyakov1}.  There
are only two cases where this relationship is well understood, 
two-dimensional Yang-Mills theory~\cite{kk,kkk,gross,gross2,gross3} 
and compact quantum
electrodynamics~\cite{polyakov2,polyakov3}.  At low temperatures 
1+1-dimensional adjoint QCD is confining in the conventional sense that
quarks only appear in the spectrum in color neutral bound states.
This is a result of the fact that, in one dimension, the gluon field
has no propagating degrees of freedom and therefore it cannot form a
color singlet bound state with an adjoint quark.  As a result, the
quantum states are color singlet bound states of two or more adjoint
quarks.  The spectrum contains an infinite number of families of
multi-quark bound states which resemble asymptotically linear Regge
trajectories~\cite{kutasov,dk,Bhanot,Bh2} and, for large energies, the
density of states increases exponentially with energy~\cite{kogan}.
This implies a Hagedorn transition~\cite{hagedorn,hgd2,hgd3} at high
temperature.  Kutasov~\cite{kutasov} supported this view by using an
argument originally due to Polchinski~\cite{polchinski} that a
deconfinement transition occurs when certain winding modes become
tachyonic at high temperature.  This behaviour was a feature of the
explicit first order deconfinement transition found in the model
considered in~\cite{stz,sz}.  That model, which coincides with the
heavy quark limit of adjoint QCD, is effectively a statistical
mechanical model for strings of electric flux, with quarks attached
to their ends.

The large-$N$ expansion of two-dimensional adjoint QCD has the same
complexity as the large-$N$ expansion of a higher dimensional
Yang-Mills theory and the leading order, infinite-$N$ limit cannot be
found analytically~\cite{thooft}.  In fact, the dimensional reduction
of three or four-dimensional Yang Mills theory produces 
two-dimensional QCD with massless adjoint scalar quarks, so the
combinatorics of planar diagrams is very similar.

\subsection{Overview}

This paper is organized as follows: In Section 2.1 we identify the
gauged principal chiral model which corresponds to the gas with 
sources in various representations. In Section
2.2 the quantum mechanical formulation of this unitary matrix
model is analyzed. This is followed by a section (2.3) where we rewrite the
model in terms of collective variables (eigenvalue density), which are
convenient for analyzing the large-$N$ limit.

Parametrized solutions to the collective field equations are given in 
Section 3.1,
and the parametrized free energy and its derivatives are obtained in 3.2.
We also show that they give rise to a first order differential equation
the free energy has to obey. In Section 3.3 we establish the existence
of a third order line in the phase diagram, and compute the point where
it terminates. Using numerical techniques we show in 3.4 that the critical
line continues from that point as a first order line. In Section 3.5 
we discuss the phase diagram we obtained. The paper ends with a 
summary (Section 4).

\section{Formalism}
\setcounter{equation}{0}
\subsection{Effective action}
The partition function of 1+1-dimensional Yang-Mills theory at
temperature $T$ and coupled to a number $K$ of non-dynamical quarks at
positions $x_1\ldots,x_K$ in representations $R_1,\ldots,R_K$ of the
gauge group is obtained by taking the thermal average of an ensemble of
Polyakov loop operators
\begin{equation}
Z[T;x_1,\ldots,x_K;R_1,\ldots,R_K]~=~\int dA_\mu
~e^{-S[A]}~\prod_{i=1}^K{\rm Tr} {\cal P}e^{i
\int_0^{1/T}d\tau A^{R_i}_0(\tau,x_i)} \; ,
\label{ftpi}
\end{equation}
where the Euclidean action is
\begin{equation}
S[A]~=~\int_0^{1/T}d\tau
\int dx\frac{1}{2e^2}{\rm
Tr}\left(F_{\mu\nu}(\tau,x)\right)^2 \; ,
\label{Seucl}
\end{equation}
the gauge fields are Hermitean $N\times N$ matrix valued vector fields
which have periodic boundary conditions in imaginary time
\[
A_\mu(\tau,x)=A_\mu(\tau+1/T,x) \; , 
\nonumber
\]
and the field strength is
\[
F_{\mu\nu} \; \equiv \; \partial_\mu A_\nu-\partial_\nu A_\mu-
i\left[ A_\mu,A_\nu \right] \; .
\nonumber
\]
The gauge field can be expanded in basis elements of the Lie algebra $A^{R_i}_\mu \equiv A^a_\mu T^a_{R_i}$ with $T^a_{R_i}$ the generators
in the representation $R_i$. For concreteness, we consider $U(N)$
gauge theory and denote the generators in the fundamental representation as
$T^a$ with $a=1,\ldots,N^2$. They obey
\begin{equation}
\left[ T^a, T^b \right] \; = \; if^{abc}T^c \; ,
\label{gen1}
\end{equation}
normalized so that
\begin{equation}
{\rm Tr} \; T^a T^b~=~\frac{1}{2}\delta^{ab} \; ,
\label{gen2}
\end{equation}
and with the sum rule
\begin{equation}
\sum_{a=1}^{N^2} T^a_{ij}T^a_{kl}~=~\frac{1}{2}\delta_{jk}\delta_{il} \; .
\label{gen3}
\end{equation}
We remark that group elements $g^{Ad}$ in the adjoint representation
are related to the fundamental representation matrices $g$ by
\begin{equation}
(g^{Ad})^{a b} \; = \; 2 \; \mbox{Tr} ( g^\dagger T^a g T^b ) \; .
\label{gadj}
\end{equation}

The expression (\ref{ftpi}) can be obtained by canonical quantization
of 1+1-dimensional Yang-Mills theory with Minkowski space action
coupled to some non-dynamical sources
\begin{equation}
S=-\int dtdx~\frac{1}{2e^2}{\rm Tr}F_{\mu\nu}F^{\mu\nu}~+~{\rm source~
terms} \; .
\label{sourceaction}
\end{equation}
In the following, we will review an argument for representing the
partition function of Yang-Mills theory as a gauged principal chiral
model which was first given in~\cite{gsst,gsst2} and which was generalized
to the case of Yang-Mills theory with sources in~\cite{stz,sz}.  As
is usual in canonical quantization of a gauge theory, the canonical 
conjugate $E(x)$ of the spatial component of the gauge field 
(which we denote by $A(x)$), is proportional to the electric field,
 $$E \; = \; \frac{1}{e^2}F_{01} \; ,$$ and obeys the commutation relation \begin{equation}
\left[ A^a(x), E^b(y) \right]~=~i\delta^{ab}\delta(x-y) \; .
\label{cancom}
\end{equation}
The Hamiltonian is
\begin{equation}
H=\int dx~ \frac{e^2}{2} \sum_{a=1}^{N^2} (E^a(x))^2 \; .
\label{ham}
\end{equation}
This Hamiltonian must be supplemented by the Gauss' law constraint
equation which is the equation of motion for $A_0$ following from
(\ref{sourceaction}) and which contains the color charge densities of
the sources
\begin{equation}
{\cal G}^a(x)\equiv \left( \frac{d}{dx}E^a(x) -
f^{abc}A^b(x)E^c(x) + \sum_{i=1}^K T^a_{R_i}\delta(x-x_i)\right)
\sim 0 \ .
\label{gauss}
\end{equation}
Here, the particles with color charges are located at positions
$x_1,\ldots ,x_K$.  $T^a_{R_i}$ are generators in the representation
$R_i$ operating on the color degrees of freedom of the {\it i}'th
particle.

There are two options for imposing this constraint. The first is to
impose another gauge fixing condition such as $$A ~\sim~0\: ,$$ and to
use the constraints to eliminate both $E$ and $A$.  The resulting
Hamiltonian is
\begin{equation}
H=\sum_{i<j,a}\frac{e^2N}{4}T^a_{R_i}\otimes T^a_{R_j}
\left| x_i-x_j\right| \; ,
\label{nambuham}
\end{equation}
which was considered in Ref.~\cite{nambu}.  It is the energy of an
infinite range spin model where the spins take values in the Lie
algebra of $U(N)$.

The other option, which makes the closest contact with string dynamics,
is to impose the constraint (\ref{gauss}) as a physical state
condition,
\[
{\cal G}^a(x)~\Psi_{\rm phys}~=~0 \; .
\]
To do this, it is most illuminating to work in the functional
Schr\"odinger picture, where the states are functionals of the gauge
field, $\psi[A]$ and the electric field is the functional derivative
operator $$ E^a(x)\Psi[A]~=~\frac{1}{i}\frac{\delta}{\delta
A^a(x)}~\Psi[A] \; , $$ The time-independent functional Schr\"odinger
equation is
\[
\int dx\left( -\frac{e^2}{2}\sum_{a=1}^{N^2}
\frac{\delta^2}{(\delta A^a(x))^2}\right)
~\Psi^{a_1\ldots a_K}\left[A;x_1, \ldots, x_K\right]~
  =~{\cal E}~\Psi^{a_1\ldots
a_K}\left[A;x_1, \ldots, x_K\right] \; .
\]
Gauss' law implies that the physical states, i.e. those which obey the
gauge constraint (\ref{gauss}), transform as
\[
\Psi^{a_1\ldots a_K}\left[ A^g;x_1, \ldots, x_K\right] \; \; = \; \;
g^{\rm R_1}_{a_1b_1}(x_1)\ldots g^{\rm R_K}_{a_Kb_K}(x_K)
\; \Psi^{b_1\ldots b_K}\left[A;x_1, \ldots, x_K\right] \; ,
\]
where $$ A^g \; \; \equiv \; \; 
gAg^{\dagger}-ig\nabla g^{\dagger} \; , $$ is the gauge transform of $A$.

For a fixed number of particles, the quantum mechanical problem is
exactly solvable.  For example, the wavefunction of a fundamental
representation quark-antiquark pair is
\begin{equation}
\Psi^{ij}[A;x_1,x_2] \; = \; \left( {\cal P}e^{i\int_{x_1}^{x_2}dyA(y)}
\right)^{ij} \; ,
\label{wf1}
\end{equation}
where the path ordered phase operator represents a string of electric
flux connecting the positions of the quark and anti-quark.  The energy
is $\frac{e^2N}{4}\vert x_1-x_2\vert$.  For a pair of adjoint
quarks, the wavefunction is
\begin{equation}
\Psi^{ab}[A;x_1,x_2] \; = \; {\rm Tr}\left( T^a{\cal P}e^{i\int_{x_1}^{x_2}A}
T^b{\cal P}e^{i\int_{x_2}^{x_1}A} \right) \; .
\label{wf2}
\end{equation}
with energy $\frac{e^2N}{2} \vert x_2-x_1\vert$.  These energy states
are identical to what would be obtained by diagonalizing the `spin'
operators in the gauge fixed Hamiltonian (\ref{nambuham}).

Note that the wavefunctions (\ref{wf1}) and (\ref{wf2}) are not
normalizable by functional integration over $A$.  This is a result of
the fact that the gauge freedom has not been entirely fixed, so that
the normalization integral still contains the infinite factor of the
volume of the group of static gauge transformations.

In general, for a fixed distribution of quarks, a state-vector is
constructed by connecting them with appropriate numbers of strings of
electric flux so that the state is gauge invariant.  The number of
ways of doing this fixes the dimension of the quantum Hilbert space.
If the flux strings overlap, the Hamiltonian can mix different
configurations, so the energy eigenstates are superpositions of string
configurations.  However, this mixing is suppressed in the large-$N$
limit (i.e. the strings are non-interacting) and any string
distribution is an eigenstate of the Hamiltonian with eigenvalue
$(e^2N/4)\times$(total length of all strings).

We shall study the thermodynamics of this system by constructing the
partition function.  We work with the grand canonical ensemble and
assume that the quarks obey Maxwell-Boltzman statistics.  The
partition function of a fixed number of quarks is constructed by
taking the trace of the Gibbs density $e^{-H/T}$ over physical states.
This can be implemented by considering set of all states in the
representation of the commutator (\ref{cancom}), spanned by, for
example, the eigenstates of $A^a(x)$ and an appropriate basis for the
quarks $$\vert A \rangle\otimes e_{a_1}\otimes
e_{a_2}\otimes\ldots\otimes e_{a_K}~~.$$ Projection onto physical,
gauge invariant states is done by gauge transforming the state at one
side of the trace and then integrating over all gauge transformations
(and then dividing by the infinite volume of the gauge group)~\cite{gpy}. 
The resulting partition function is
\begin{equation}
Z[T/e^2;x_1,\ldots,x_K] \; = \; \int[dA][dg]~\left<A\right| e^{-H/T}
\left|A^g\right> {\rm Tr}~g^{ R_1}(x_1)\ldots {\rm Tr}~g^{ R_K}(x_K)
~~,
\label{hkpi}
\end{equation}
where $[dg]$ is the Haar measure on the space of mappings from the
line to the group manifold and $[dA]$ is a measure on the convex
Euclidean space of gauge field configurations.  The expression
(\ref{hkpi}) is identical to (\ref{ftpi}) with the Polyakov loop
operator is the trace of the group element $g(x)$ in the appropriate
representation.

In going over to the grand canonical ensemble the first step is to
integrate over all particle positions. We then multiply by the
fugacities for each type of charge: a factor of $\lambda_R$ for each
quark in representation $R$.  To impose Maxwell-Boltzmann statistics,
we divide by the factorial of the number of quarks in each
representation.  We then sum over all numbers of quarks in each
representation.  This exponentiates the fugacities, resulting in the
grand partition function
\begin{equation}
Z[T/e^2,\lambda_R]~=~\int [dA][dg]~e^{-S_{\rm eff}[A,g]} \; ,
\label{partition}
\end{equation}
where the effective action is
\begin{equation}
\exp \Big( -S_{\rm eff}[A,g] \Big) = \left< A\right| e^{-H/T}\left| A^g
\right>\exp \left(\int dx~\sum_{R}\lambda_R {\rm Tr}~ g^R(x) \right) \; ,
\label{Seff0}
\end{equation}
and the summation in the exponent is over all the irreducible
representations of $U(N)$ we want to consider.  
The Hamiltonian is the Laplacian on the
space of gauge fields.  The heat kernel obeys the equation
\[
\left( T^2\frac{\partial}{\partial T}+\int dx\frac{e^2}{2}
\sum_{a=1}^{N^2} \left( \frac{\delta}{\delta A^a(x)} \right)^2
\right)\left< A\right|
e^{-H/T}\left| A^g\right> \; = \; 0 \; ,
\]
with the boundary condition
\[
\lim_{1/T\rightarrow0}\left< A\right| e^{-H/T}\left|A^g\right>
\; \; = \; \; \delta(A-A^g) \; .
\]
These equations are easily solved by a Gaussian - divided by a
$T$-dependent constant: $$
\left< A\right| e^{-H/T} \left|
A^g\right> \; \; \sim \; \;
\exp\left(-\int dx ~\frac{T}{e^2}~{\rm Tr}~(A-A^g)^2\right)\ .
$$

We see that the effective theory is the gauged principal chiral model
with a potential energy term for the group-valued degrees of freedom,
\begin{equation}
S_{\rm eff}[A,g]~=~ \int dx~\left( \frac{N}{2 \gamma}~ {\rm Tr}\Big|
\nabla g(x) + i[A(x),g(x)]\Big|^2 -\sum_{R}\left(
\lambda_R
{\rm Tr}~g^R(x)  \right)\right) \label{seff} \; .
\label{seffrep}
\end{equation}
Note that we have introduced the coupling constant
\begin{equation}
\gamma \; \; \equiv \; \; \frac{ e^2 \; N}{2 T} \; .
\label{gammadefn}
\end{equation}
When we analyze the limit $N \rightarrow\infty$, we will tune $e^2$
such that $\gamma$ is constant. Moreover, we assume that 
the fugacities $\lambda_R$
are scaled such that all terms in the action (\ref{seffrep}) are of order
$N^2$.

The potential energy term in the effective action,
\begin{equation}
V(g)~\equiv~-\sum_R \lambda_R {\rm Tr}\left(g^R(x)\right) \; ,
\label{poten}
\end{equation}
is the expansion of a local class function of the group element $g(x)$
(one which obeys $V(g)=V(hgh^{-1})$ for $h\in U(N)$) in group
characters with coefficients $\lambda_R$.  The characters
\[
\chi_R(g)~\equiv~{\rm Tr}\left( g^R(x) \right) \; ,
\]
form a complete set of orthonormal class functions of the group
variable, with inner product
\[
\int [dg]\chi_R^*(g)\chi_{R'}(g)~=~\delta_{R,R'} \; .
\]
Here [dg] is not a functional integral measure, but is the Haar
measure for integration on $U(N)$. From the potential, we can find a
fugacity by
\[
\lambda_R~=~-\int[dg]\chi_R^*(g)V(g) \; .
\]
By tuning the fugacities appropriately, we could obtain any local
invariant potential.

The effective action (\ref{seff}) with all $\lambda_R=0$ was discussed
by Grignani et.al.~\cite{gsst,gsst2} and was solved explicitly in the limit
$N\rightarrow \infty$ by Zarembo~\cite{Zar,Zar2}. The model with
$\lambda_{Ad}\neq 0$ (with adjoint quarks) was solved in
Refs.~\cite{stz} and~\cite{sz}.  The effective action (\ref{seff}) is
gauge invariant, $$S_{\rm eff}[A,g]~=~ S_{\rm eff}[A^h,
hgh^{\dagger}]~~. $$
It is also covariant under the global transformation
\begin{equation}
S_{\rm eff}[A,zg,\lambda_R]~=~ S_{\rm eff}[A,g,z^{-C_1(R)}\lambda_R] \; .
\label{zsym}
\end{equation}
where $z$ is a constant element from the center of the gauge group,
which for $U(N)$ is $\sim$ $U(1)$ and would be the discrete group
$Z_N$ for gauge group $SU(N)$. Here, $C_1(R)$ is the linear Casimir
invariant of the representation $R$, which is the number of boxes in
the Young tableau corresponding to $R$.  When the gauge group is $SU(N)$
and the only non-zero fugacities are for the zero `N-ality'
representations, i.e. those for which $C_1(R)=0$ mod $N$, there is a
global $Z_N$ symmetry.  For gauge group $U(N)$, this occurs only when
all representations with non-zero fugacities have equal numbers of
quarks and anti-quarks.  The fugacities of other non-symmetric
charges, can be thought of as an external field which breaks the
center symmetry of the system explicitly. This situation is akin to
the effect of an external magnetic field on a spin system.

\subsection{Matrix quantum mechanics}

If we re-interpret $x$ as Euclidean time, the partition function that
we have derived has the form of a Euclidean space representation of
the partition function for matrix quantum mechanics, where the free
energy is identical to the ground state energy of the matrix quantum
mechanics.  We can study the latter model by mapping the problem to
real time $\tau$ by setting $x = i\tau$ and $A\rightarrow -iA$.  The
action in real time is then
\[
S_{QM} \; = \;
\int d\tau \left( \frac{N}{2\gamma}{\rm Tr}\left|\dot g+i\left[
A,g\right]\right|^2 -V(g)\right) \; .
\]
We remark that this action must not be confused with the action
(\ref{Seucl}). $S_{QM}$ is the action for a 0+1-dimensional problem
(quantum mechanics), while (\ref{Seucl}) is the action for Yang Mills
theory in 1+1 dimensions. This remark also holds for the Hamiltonian
below. In order to avoid confusion, we label the quantum mechanical
quantities with the subscript $QM$.

The canonical momentum conjugate to the group valued position variable
$g$ is the Lie algebra element
\[
\Pi \; = \; 
\frac{N}{\gamma}\left( ig^{\dagger}\dot g +g^{\dagger}Ag-A\right) \; ,
\]
and the Hamiltonian is
\begin{equation}
H_{QM} \; = \; \frac{\gamma}{2N}{\rm Tr}\Pi^2-\frac{\gamma}{N}{\rm
Tr}\Pi(g^{\dagger}Ag-A) ~+~V(g) \; .
\label{Cqmham}
\end{equation}
The gauge field $A$ plays the role of a Lagrange multiplier which
enforces the constraint
\[
g\Pi g^{\dagger}-\Pi\sim0 \; ,
\]
and the Hamiltonian reduces to
\[
H_{QM} \;  = \; \frac{\gamma}{2N}{\rm Tr}\Pi^2 +V(g) \; .
\]
We can expand the canonical momentum as
\[
\Pi~=~ \sum_a \Pi^a T^a \; ,
\]
Then, the components satisfy the Lie algebra
\begin{eqnarray}
\left[ \Pi^a, \Pi^b\right] \; &=& \; if^{abc}\Pi^c \; , \\
\left[ \Pi^a,g \right] \; &=& \; gT^a \; , \\
\left[ \Pi^a, g^{\dagger} \right] \; &=& \;  -T^a g^{\dagger} \; .
\end{eqnarray}
It follows that in the Schr\"odinger picture the components of
the canonical momentum are represented as
\[
\Pi^a~=~{\rm Tr}gT^a\frac{\partial}{\partial g}~=~g_{ij}T^a_{jk}
\frac{\partial}{\partial g_{ik}} \; .
\]
Denoted in components the constraint reads
\[
G^a \; \equiv \; \left( {\rm Tr}T^a gT^b g^{\dagger}-
\frac{1}{2}\delta^{ab} \right)
\Pi^b \sim0 \; .
\]
The constraint has no operator ordering ambiguity.  It generates the
adjoint action of the symmetry group
\[
\left[ G^a,g\right] \; = \; \frac{1}{2} \left[ T^a, g\right] \; .
\]
The constraint can be realized as a physical state condition
\[
G^a~\psi_{\rm phys}~=~0 \; .
\]
In the representation where states are functions of $g$, this implies
that the physical states are class functions
\[
\psi_{\rm phys}(g)~=~\psi_{\rm phys}(hgh^{-1}) \; ,
\]
where $h\in U(N)$. This means that the physical states are functions
of the eigenvalues of $g$.  In a basis where $g$ is diagonal,
\begin{equation}
g~=~\left( \matrix{ e^{i\alpha_1} &0& 0 & 0&\ldots &0\cr 0
&e^{i\alpha_2}&0&0&\ldots&0\cr 0&0& e^{i\alpha_3}&0&\ldots&0\cr
0&0&\ldots & & & 0\cr 0&0&\ldots & & & e^{i\alpha_N}\cr }\right) \; ,
\label{diagg}
\end{equation}
the wavefunctions are functions of $\alpha_i$,
\begin{equation}
\psi_{\rm phys}(\alpha_1,\ldots,\alpha_N)=\psi_{\rm phys}(\alpha_1,
\ldots,\alpha_i+2\pi,\alpha_N) \; .
\label{psialph}
\end{equation}
Denoting the gauge group Laplacian in components
\begin{equation}
\triangle \; \; \equiv \; \; \sum_{a=1}^{N^2} ( \Pi^a )^2 \; ,
\label{glap}
\end{equation}
the Hamiltonian reads
\[
H_{QM} \; \; = \; \; \frac{\gamma}{4N}\Delta \; + \; V(g) \; .
\]
Since the potential $V(g)$ is also a class function and depends only
on the eigenvalues, when operating on the physical states, the
Hamiltionian can be expressed in terms of eigenvalues and derivatives
by eigenvalues
\[
H_{QM} \; =
\; \frac{\gamma}{4N}\frac{1}{\tilde J(\alpha)}\left(\sum_1^N
-\frac{\partial^2} {\partial\alpha_i^2}-N(N^2-1)/12\right)
\tilde J(\alpha)+V(\alpha) \; ,
\]
where
\[
\tilde J(\alpha)=\prod_{i<j} 2\sin\frac{1}{2}(\alpha_i-\alpha_j)
=\frac{1}{(2i)^{N(N-1)/2}}\frac{ J(\alpha)}{\prod_i z_i^{(N-1)/2} } \; ,
\]
and
\[
J(z)=\prod_{i<j}(z_i-z_j)~~~,~~z_i\equiv e^{i\alpha_i} \; .
\]

The physical states must by symmetric functions of $\alpha_i$. (There
is a residual gauge invariance~\cite{lang,lang2} under the Weyl group which
permutes the eigenvalues and the physical state condition requires
that the physical states be symmetric under these permutations.)
The normalization integral for the wavefunction is
\[
\int [dg] \psi^{\dagger}(g)\psi(g)~=~1 \; .
\]
Since the integrand depends only on the eigenvalues of $g$, It is
convenient to write the Haar measure as an integral over eigenvalues
of $g$ with the Jacobian which is the Vandermonde determinant,
\[
\int (\prod_i d\alpha_i) \vert \tilde J(\alpha)\vert^2
\psi^{\dagger}(\alpha)
\psi(\alpha)~=~1 \; .
\]
The Hamiltonian and inner product have a particularly simple form when
we redefine the wavefunction as
\[
\tilde\psi(\alpha_1,\ldots,\alpha_N)\equiv \tilde J(\alpha) \psi(\alpha_1,
\ldots,\alpha_N) \; .
\]
Since $\tilde J$ is antisymmetric, $\tilde\psi$ is a completely
antisymmetric function of the eigenvalues, which we can think of as
the coordinates of fermions.  The Hamiltonian is that of an
interacting Fermi gas
\[
\left\{
\frac{\gamma}{4N}\left( \sum_1^N -\frac{\partial^2}{\partial\alpha_i^2}
-N(N^2-1)/12\right)+V(\alpha)\right\}\tilde\psi(\alpha)={\cal E}\tilde
\psi(\alpha) \; .
\]
This correspondence of a c=1 matrix model with a Fermi gas was first
pointed out in Ref.~\cite{bipz}.
\subsection{Large N: Collective variables}

In this section we shall examine the collective field formulation of the
large-$N$ limit of the theory that we discussed in the last 
subsection~\cite{Zar,JS,JS2,JS3}. 
The Hamiltonian obtained in the last subsection reads
\begin{equation}
H_{QM}  \; \; = \; \; \frac{\gamma}{4 N} 
\sum_{a=1}^{N^2} ( \Pi^a )^2
 \; + \; V(g) \; ,
\label{ham2}
\end{equation}
with (compare (\ref{poten}))
\[
V(g)~\equiv~-\sum_R \lambda_R {\rm Tr}\left(g^R(\tau)\right) \; .
\]
It was shown (compare (\ref{diagg}),(\ref{psialph}))
that the wavefunction depends only on the eigenvalues $e^{i\alpha_j}$
of $g$ and thus the density of eigenvalues
\[
\rho(\theta,\tau)~\equiv~\frac{1}{N}\sum_{i=1}^N
\delta(\theta-\alpha_i(\tau)) \; ,
\]
completely characterizes the properties of the system.
Interpretation of the physics of the system at large $N$
is more convenient
when one considers the Fourier transform of the eigenvalue distribution
\begin{equation}
\rho(\theta,\tau)~=~\frac{1}{2\pi}~+~
\frac{1}{2\pi}
\sum_{n \neq 0} c_n(\tau) e^{-in\theta} \; ,
\label{den}
\end{equation}
where we have defined
\[
c_n(\tau) \; \equiv \; \frac{1}{ N} \Tr g^n(\tau) ~~~~,~~~
c_{-n}(\tau) \; = \; \overline{c_n(\tau)} \; .
\]
We now turn our attention to developing the collective field theory
formulation of the $\mbox{(thermo-)}$ dynamical problem given by the
Hamiltonian (\ref{ham2}).  Since the wavefunction depends only on
the eigenvalues of $g$, we would like a Hamiltonian
equivalent to (\ref{ham2}) but written in terms of
the eigenvalue density $\rho$ and a conjugate momentum $\Pi$.
At large $N$ we will find this Hamiltonian and write equations
of motion for $\rho$ and $\Pi$. So far we have not imposed any restriction 
on the potential $V(g)$, but from now on we assume, that it can be 
expressed as a functional of the eigenvalue density $\rho(\theta)$. 
In particular 
the potential we are going to analyze below will have this property.

The canonical momentum operates on the wavefunction as
\[
\Pi^a\psi[\rho]=\int d\theta\left[\Pi^a,\rho(\theta)\right] \frac{\delta}
{\delta\rho(\theta)}~\psi[\rho]
\]
\[
=\frac{1}{2\pi N}\int d\theta\sum_K e^{-iK\theta}K{\rm Tr}(T^ag^K)
\frac{\delta}{\delta\rho(\theta)}\psi[\rho] \; ,
\]
and the Laplacian (\ref{glap}) is
\begin{eqnarray}
\Delta\psi[\rho]~=~\left(
\frac{1}{4\pi N}\int d\theta\sum_K e^{-iK\theta}\left| K\right|\left(
\sum_{L=0}^K {\rm Tr}g^L {\rm Tr}g^{K-L}-N{\rm Tr}g^K\right)
\frac{\delta}{\delta\rho(\theta)}
\right.\nonumber\\ \left.
+ \; \frac{1}{8\pi^2N^2}\int d\theta d\theta'
\sum_{KL}K L e^{-iK\theta-iL\theta'} {\rm
Tr}g^{K+L}~\frac{\delta^2}{\delta\rho(\theta)\delta\rho(\theta')}
\right)~\psi[\rho] \; ,
\nonumber
\end{eqnarray}
which can be written as
\[
\Delta\psi[\rho] =
-\frac{1}{2N}\int d\theta\rho(\theta)\Bigg\{\left( \frac{\partial}
{\partial\theta}\frac{\delta}{\delta\rho(\theta)}\right)^2 
\]
\[
- N^2{\cal P}\int d\theta' \rho(\theta')\cot\left(
\frac{\theta-\theta'}{2}\right)\frac{\partial}{\partial\theta}
\frac{\delta}{\delta\rho(\theta)}
\Bigg\}\psi[\rho] =
\]
\[
-\frac{1}{2N}\int d\theta\rho(\theta)\left(\left(
\frac{\partial}{\partial\theta}
\frac{\delta}{\delta\rho(\theta)}+{\cal V}(\theta)\right)^2 -
{\cal V}^2(\theta)\right)\psi[\rho] \; ,
\]
where
\[
{\cal V}(\theta)=\frac{N^2}{2}{\cal P}\int d\theta'\rho(\theta')\cot\left(
\frac{\theta-\theta'}{2}\right) \; .
\]
${\cal P}$ indicates principal value integral.

The transformation of the wavefunction
\[
\psi[\rho] \; = \; \tilde\psi[\rho]\exp\left(-\frac{N^2}{2}\int
d\theta d\theta'
\ln\sin\frac{\left|\theta-\theta'\right|}{2} \rho(\theta) \rho(\theta')
\right) \; ,
\]
transforms the derivative in the Schr\"odinger equation so that it has
the form
\begin{equation}
\left\{-\frac{\gamma}{8N^2}\int d\theta \rho(\theta)\left\{ \left(
\frac{\partial}{\partial\theta}\frac{\delta}{\delta\rho(\theta)}
\right)^2-{\cal V}^2(\theta)\right\}+V[\rho]\right\}~\tilde\psi[\rho]
~=~E~\tilde\psi[\rho] \; .
\label{jsham}
\end{equation}
The second term on the left-hand-side of this equation has a simple
form.  In~\cite{GaPaSe97} it is shown that it gives rise 
to a term which is cubic in the density.
Thus, the Schr\"odinger equation has the form
\begin{equation}
\left\{-\frac{\gamma}{8N^2}\int d\theta \left(\rho(\theta) \left(
\frac{\partial}{\partial\theta}\frac{\delta}{\delta\rho(\theta)}
\right)^2-N^4\frac{\pi^2}{3}\rho^3(\theta)
\right)+V[\rho]\right\}~\tilde\psi[\rho]~
=~E~\tilde\psi[\rho] \; ,
\label{jsham2}
\end{equation}
up to an overall constant.

The large-$N$ limit is dominated by the eikonal approximation. In this
approximation, we make the ansatz for the wavefunction
\[
\tilde\psi[\rho]~=~\exp\left( iN^2 S[\rho]\right) \; .
\]
The eikonal, $S$ then obeys the equation
\begin{equation}
\frac{H_{QM}[\rho,\Pi]}{N^2}~=~
\frac{\gamma}{8}\int d\theta \left\{\rho(\theta) \left(
\frac{\partial }{\partial\theta}\frac{\delta~S}{\delta\rho(\theta)}
\right)^2+\frac{\pi^2}{3}\rho^3(\theta)
\right\}+\frac{1}{N^2}V[\rho] ~=~\frac{E}{N^2} \; .
\label{schroedeqn}
\end{equation}
Here, we have ignored a term which is of subleading order in $N^2$.
We have also assumed that $V[\rho]$ will be of order $N^2$ (compare Section
2.1) and that the
natural magnitude of the energy eigenvalue is of order $N^2$.

To solve this equation for the ground state, we must find its minimum
by varying $\rho$ and the canonical momentum
$$
\Pi \; = \; \delta S/\delta\rho \; ,
$$
subject to the condition that $\rho$ is normalized.  This leads to
the equations of collective field theory
\begin{eqnarray}
\frac{\partial}{\partial \tau}
\rho(\tau,\theta) \; \; &=&  \; \;
\frac{\; \delta H_{QM}/N^2 \; }{\delta \Pi(\tau,\theta)} \; ,
\nonumber\\
\frac{\partial}{\partial \tau}
v(\tau,\theta) \; \; &=& \; \; -\frac{\partial}{\partial\theta} \;
\frac{ \; \delta H_{QM}/N^2 \; }{\delta \rho(\tau,\theta)} \; ,
\nonumber
\end{eqnarray}
where
\bd
v(\tau,\theta)~\equiv~\frac{\partial}{\partial\theta} \; 
\Pi(\tau,\theta) \; .
\ed
Taking the derivative of the second equation with respect
to $\theta$ eliminates a Lagrange
multiplier which must be introduced on order to enforce the
normalization condition for $\rho$. Using (\ref{schroedeqn}) one finds
\begin{eqnarray}
\frac{\partial}{\partial \tau}\rho \; + \; \frac{\gamma}{4}\frac{\partial}
{\partial\theta}\left(\rho v\right) \; &=& \; 0 \; ,
\nonumber\\
\frac{\partial}{\partial \tau}v \; + \; \frac{\gamma}{8}\frac{\partial}
{\partial\theta}\left( v^2+\pi^2\rho^2\right) \; + \; \frac{1}{N^2}
\frac{\partial}{\partial\theta}\frac{\delta}{\delta\rho}V[\rho] \;
&=& \; 0 \; .
\label{colfe}
\end{eqnarray}
It is interesting to note that these are nothing but Euler's
equations for a fluid with equation of state
$P= \pi^2 \rho^3 /3$ on a cylinder with coordinates
$(\theta, \tau)$. The inclusion of a potential $V(\theta,\tau)$
corresponding to non-Abelian charges is equivalent to subjecting
the fluid to an external force which is derived from $V(\theta,\tau)$.

We shall use these equations in the next section where we analyze
the large-$N$ limit of a mixed gas of adjoint and fundamental
charges.

\section{Free energy and critical behaviour}
\setcounter{equation}{0}
\subsection{Static solutions to the collective field equations}

In this section we will find static solutions to
the collective field equations (\ref{colfe}). The most simple 
potentials involve only the lowest representations, the fundamental, 
its conjugate and the adjoint. We shall consider a slight generalization 
of these and  use powers of the lowest representations to include
multiple windings of the Polyakov loop operator. Our potential 
reads
\begin{equation}
V(g)\; \equiv \; -\sum_{n=1}^\infty \left( \kappa_n N{\rm Tr}(g^n)+
\bar\kappa_n N
{\rm Tr}((g^{\dagger})^n)+\lambda_n\vert{\rm Tr}g^n\vert^2 \right) \; ,
\label{potential}
\end{equation}
where we made use of (compare (\ref{gen3}), (\ref{gadj}))
\[ 
{\rm Tr}~(g^{Ad}(x))^n \; = \;
\left| {\rm Tr}~ g^n(x)\right|^2 \; ,
\]
to relate the trace in the adjoint representation to the trace in
the fundamental representation. The couplings for the 
fundamental representation charges (and their conjugates) were chosen to 
scale $\sim N$, to make the potential of order $N^2$. It should be remarked 
that parts of this section can easily be extended to more general potentials.

The potential (\ref{potential}) indeed can be expressed as
a functional of the eigenvalue density (\ref{den}). 
The collective field Hamiltonian (\ref{schroedeqn}) then reads
\begin{equation}
\frac{H_{QM}}{N^2}~
=\frac{\gamma}{8} \int d\theta\left[ \rho(\theta)\left(
v(\theta) \right)^2
 +\frac{\pi^2}{3}\rho^3(\theta) \right] \label{col} 
\end{equation}
\[
- \sum_{n=1}^\infty
\left( \lambda_n\left| \int d\theta~\rho(\theta)e^{in\theta}\right|^2
+ \kappa_n \int~d \theta~ \rho(\theta) \e^{i n \theta} + \bar{\kappa}_n
\int ~d \theta ~\rho( \theta) \e^{-i n \theta} \right) - 
\frac{\gamma}{96}\; .
\]
In order to maintain correspondence with the original version of the
Hamiltonian (\ref{ham2}) we subtract the constant $\gamma/96$. It sets the 
energy scale such that the free energy vanishes in the confined phase 
of the model with only adjoint charges (see below). 

The corresponding collective field equations (\ref{colfe}) read
\begin{equation}
\frac{\partial\rho}{\partial x} \; + \;
\frac{\gamma}{4}\frac{\partial}{\partial\theta}
(\rho v) \; = \; 0 \; ,
\label{coll1}
\end{equation}
\begin{equation}
\frac{\partial v}{\partial x} \; + \; \frac{ \gamma}{8} \frac{\partial v^2}
{\partial\theta}
\; - \; \frac{ \pi^2\gamma}{8} \frac{\partial\rho^2}{\partial\theta} \; +
\; \frac{\partial}{\partial \theta} \sum_n \left[
(\lambda_n c_{-n} + \kappa_n) \e^{i n \theta} +
( \lambda_n c_n + \bar{\kappa}_n ) \e^{-i n \theta} \right]~=~0 \; .
\label{coll2}
\end{equation}
where $c_n $ are the $x$-dependent Fourier coefficients of $\rho$
as introduced in (\ref{den}).  Note that we also performed the change of
variables, $\tau \rightarrow -ix$ and
$v\rightarrow iv$ in these equations in order to invert
the Wick rotation performed at the beginning of Section 2.2 prior to
canonical quantization.

We will only consider real, static solutions of the non-linear
equations (\ref{coll2}), that is, where 
$\rho(\theta,x) = \rho_0(\theta)$ and the velocity $v$
vanishes identically.  Consequently
\vskip1mm
\begin{equation}
\rho_0(\theta)=\left\{ \matrix{
2 \sqrt{\frac{2}{\gamma\pi^2} }\sqrt{ E+\sum(\lambda_n c_{-n}+\kappa_n)
e^{in\theta}+\sum(\lambda_n c_n+ \bar{\kappa}_n )e^{-in\theta} } &
\mbox{where } \rho \mbox{ is real} \cr 0 & {\rm otherwise}\cr} . \right. 
\label{Cdens}
\end{equation}
The constant of integration $E$ has physical interpretation as the
 Fermi energy of a collection of $N$ fermions~\cite{bipz}
in the potential $V[\rho]$ and is fixed by the normalization condition
\beq
1 \; = \; \int d \theta ~\rho_0(\theta) \; .
\label{normcond}
\eeq
Here it is more convenient to express the $c_n$ in terms of $\rho_0$
(compare (\ref{den}))
\begin{equation}
c_n \; = \;  \int d \theta ~\rho_0(\theta) e^{in\theta} \; .
\label{cnphi}
\end{equation}
The real support of the function $\rho_0(\theta)$ is the positive support of 
$\Lambda \equiv E+\sum(\lambda_n c_{-n}+\kappa_n)
e^{in\theta}+\sum(\lambda_n c_n+ \bar{\kappa}_n )e^{-in\theta}$.  The
zeros of $\Lambda$ define the edges of the eigenvalue distribution and
when these zeros condense one has critical behaviour in the observables
of the model as in general Hermitean and unitary matrix models.

\subsection{A differential equation for the free energy}
In this subsection we compute all first order derivatives of the
free energy and show that they obey a differential equation of the
Clairaut type.

Inserting the static solution (\ref{Cdens}) in (\ref{col}) we obtain for the
free energy
\begin{equation}
\frac{1}{N^2} \langle H_{QM} \rangle \; \; \equiv \; \;
f \; \; = \; \; \frac{1}{3} E \; - \; \frac{1}{3} \sum_{n=1}^\infty
\Big[ \lambda_n |c_n|^2 \; + \; 2(\; \kappa_n c_n \; + \;
\bar{\kappa}_n c_{-n} \; ) \Big] - \frac{\gamma}{96} \; .
\label{CfreeE}
\end{equation}
Note that $f$ is the leading coefficient ($O(N^2)$)
of the energy for the matrix quantum mechanics problem, but for the
quark gas problem plays the role of the leading coefficient of the
energy {\it density}.

Deriving the expression (\ref{CfreeE}) with respect to $\lambda_J$ for
some fixed $J$ and using derivatives of Equations (\ref{normcond})
and (\ref{cnphi}) with respect to the same parameter one obtains
\begin{equation}
\frac{d f}{d \lambda_J} \; = \; -|c_J|^2 \; .
\label{Cdfdl}
\end{equation}
We use the notation $d/d\lambda_J$ to indicate that also the
$c_n$ and $E$ which implicitly depend on $\lambda_J$ are derived
with respect to this coupling. Similarly one can show
\begin{equation}
\frac{df}{d \kappa_J} \; = \; - c_J\; ,
\label{Cdfdk}
\end{equation}
and
\begin{equation}
\frac{df}{d\gamma} \; = \; \frac{1}{3 \gamma} E \; + \;
\frac{1}{3 \gamma} \sum_{n=1}^\infty \Big[ 2 \lambda_n |c_n|^2 \; +
\; \kappa_n c_n \; + \;
\bar{\kappa}_n c_{-n} \Big] \; - \; \frac{1}{96} \; .
\label{Cdfdg}
\end{equation}
It is interesting to notice, that combining Equations (\ref{Cdfdl}) -
(\ref{Cdfdg}) gives rise to a first order differential equation
of the Clairaut type
\begin{equation}
\gamma \frac{df}{d\gamma} \; + \; \sum_{n=1}^\infty
\Big[ \lambda_n \frac{df}{d\lambda_n} +
\kappa_n \frac{df}{d \kappa_n} + \bar{\kappa}_n
\frac{df}{d \bar{\kappa}_n} \Big] \; \; = \; \; f \; .
\label{Cdiffe}
\end{equation}
This differential equation has general solutions of the form
\[
f ( \gamma, \lambda_n, \kappa_n ) \; \; = \; \;
\gamma \; F( \frac{\lambda_n}{\gamma}, \frac{\kappa_n}{\gamma} ) \; ,
\]
where $F$ is some arbitrary smooth function. This result shows that
the parameter $\gamma$ is not driving the physical properties of the model,
but rather sets the energy scale. The differential equation (\ref{Cdiffe})
gives no further restrictions on the function $F$ and another
analysis will be adopted in the next section. However, when finding the
physical interpretation of the phase diagram the differential
equation (\ref{Cdiffe}) is a valuable tool.

\subsection{Regime of the third order phase transition}

We now restrict ourselves  to the case of only one pair of non-vanishing
couplings
$\lambda_J, \kappa_J \neq 0$. Furthermore it is sufficient to consider
$\kappa_J$ real, since an eventual phase of $\kappa_J$ can always be
removed by using the covariance (\ref{zsym}) of the action and
Haar measure $[dg]$ under transformations by a constant element of $U(1)$.

In the form of (\ref{Cdens}) it is evident we need to solve
simultaneously for  the normalization
condition (\ref{normcond}) and the Fourier moment (\ref{cnphi})
in order to
have a self-consistent solution of the saddle-point equations.
We begin  by introducing an auxiliary complex
parameter
\begin{equation}
r_J~e^{iJ \beta_J}\; \equiv \; \lambda_J ~ c_{-J} + \kappa_J \; ,
\label{Cname}
\end{equation}
and rescaling the Fermi energy as
\begin{equation}
E \; \; \equiv \; \;  2 \mu \; r_J \; .
\label{CEscal}
\end{equation}
With this notation the normalization and moment equations are respectively
\begin{equation}
1 \;  = \; 2\sqrt{\frac{r_J}{\gamma}} \; I(\mu)
\; \; \; \; \; \; , \; \; \; \; \; \;
c_Je^{i J \beta} \; = \; 2\sqrt{\frac{r_J}{\gamma}} \; M(\mu) \; ,
\label{Cnorm}
\end{equation}
where we have defined the integrals
\begin{eqnarray}
I(\mu) \; &\equiv& \; \frac{2}{\pi} \int_{-\pi}^{\pi} \sqrt{\mu +
\cos(J\theta)} \;
\mbox{H}(\mu + \cos(J\theta)) \; d\theta \; ,\nonumber \\
M(\mu) \; &\equiv& \; \frac{2}{\pi} \int_{-\pi}^{\pi} \cos{J\theta}
\sqrt{\mu + \cos(J\theta)} \;
\mbox{H}(\mu + \cos(J\theta)) \; d\theta \; .
\label{Cparint}
\end{eqnarray}
Here H(..) denotes the step function. A simple transformation of the
integration variable shows that $I(\mu)$ and $M(\mu)$ are independent of
$J$. Thus $J$ enters only as the subscript of the parameters. For
notational convenience we abbreviate
\begin{equation}
\lambda_J \; \equiv \; \lambda \; \; \; \; , \; \; \; \;
\kappa_J e^{-iJ\beta_J} \; \equiv \; \kappa \; .
\label{abbr}
\end{equation}
We remark that $I(\mu), M(\mu)$,
and thus $c_J e^{iJ\beta_J}$ are real. 
Eliminating the moment $c_J$ from
(\ref{Cnorm}) by using the definition (\ref{Cname}) we obtain
\begin{equation}
\frac{{\kappa}}{\gamma} \; \; = \; \; \frac{1}{I(\mu)^2}
\left[ \frac{1}{4} - \frac{\lambda}{\gamma} I(\mu) M(\mu) \right] \; .
\label{Cnecc}
\end{equation}
This family of lines in the $\kappa, \lambda$-plane parametrized
by $\mu$
represent a necessary condition which a solution of the normalization and
moment equations (\ref{Cnorm}) must obey. From the last equation it is 
obvious that also the product $
\kappa \; = \; \kappa_J e^{-iJ\beta} $
is real (thus $e^{iJ\beta}$ is just a sign). It occurs as a natural
parameter when rewriting the free energy in terms of $I(\mu)$ and $M(\mu)$
(use (\ref{CfreeE}))
\begin{equation}
f \; = \; \frac{\gamma}{12 I(\mu)^2} \left[
2\mu - \frac{M(\mu)}{I(\mu)} \right] \; - \;
\kappa \frac{M(\mu)}{I(\mu)} \; - \frac{\gamma}{96} \; ,
\label{Cfmu}
\end{equation}
where we have eliminated $\lambda$ using the necessary
condition (\ref{Cnecc}). Also the first derivative of the
free energy with respect to $\kappa$ can be expressed conveniently in terms of $I(\mu)$ and $M(\mu)$
\begin{equation}
\frac{d  f}{d \kappa } \; = \;  -2 \frac{M(\mu)}{I(\mu)} \; .
\label{Cdfmu}
\end{equation}
Remember that we restricted ourselves to $\kappa_J$ real, and thus we
encounter a factor 2 compared to \ref{Cdfdk}), since a real $\kappa_J$
is the same for both terms $c_J$ and $c_{-J}$ in the potential
(\ref{CfreeE}).

With the parametric solution (\ref{CfreeE}) at hand we turn our attention
to establishing the critical behaviour in this model.
In~\cite{GaPaSe97} it is shown that the first derivatives of $I(\mu)$ and
$M(\mu)$ have non-analytic behaviour at $\mu = 1$, hence the expression
(\ref{Cdfmu}) suggests that the vicinity of $\mu = 1$ is a
natural place to
look for non-analytic behaviour in the free energy of our model.
Using the explicit results for $I(1), M(1)$ (see~\cite{GaPaSe97}) and
(\ref{Cnecc}) we obtain the necessary condition for the critical
($\mu = 1$) values of $\lambda$ and $\kappa$
\begin{equation}
\frac{\kappa^c}{\gamma} \; \; = \; \;  \frac{\pi^2}{512} \; -
\; \frac{1}{3}\frac{\lambda^c}{\gamma}  \; .
\label{Ccritlin}
\end{equation}
Having identified a line in the phase space where we expect critical behaviour
we will now proceed to establish the details of this critical behaviour.
Following~\cite{gubser,gubser2,gubser3}
we begin by expanding about $\mu =1$ and the line
(\ref{Ccritlin})
\begin{equation}
\mu = 1 + \varepsilon, \varepsilon > 0 \; \; \; \; ,  \; \; \; \;
I(1 + \varepsilon ) = I_c + \delta I \; \; \; \; , \; \; \; \;
M(1 + \varepsilon ) = M_c + \delta M \; ,
\label{Cexp}
\end{equation}
($I_c \equiv I(1), M_c \equiv M(1)$) and analyze the variation of $\kappa$
around $\kappa^c$ while keeping $\lambda$ fixed
\begin{equation}
\kappa = \kappa^c + \delta \kappa
\; \; \; \; ,
\; \; \; \; \lambda = \lambda^c \; .
\label{Cexp2}
\end{equation}
Due to the nontrivial support of the integrands in (\ref{Cparint})
in principle one has to distinguish the cases $\mu > 1$ and $\mu < 1$;
(see~\cite{GaPaSe97} for details). In order to keep the formulas simple,
we explicitly analyze only the case $\mu > 1$ as given by
(\ref{Cexp}), (\ref{Cexp2}). The case $\mu < 1$ can be treated along the
same lines and we denote the corresponding results in the end.

The expansion now consists of two steps. We first
expand the necessary condition
(\ref{Cnecc}) at $\mu =1$ to obtain the relation between the variation
$\delta \kappa $ and $\varepsilon$. In the second step we
expand the right hand side of (\ref{Cdfmu}) at $\mu = 1$ and use the
result of step one to express the variation of $ df/d\kappa$
in terms of $\delta \kappa $. The latter result can then be
used to analyze eventual singular behaviour of higher derivatives of the
free energy.

Expanding the necessary condition (\ref{Cnecc}) and using (\ref{Ccritlin})
we obtain for the variation of $\kappa$ to lowest order
\begin{equation}
\frac{\delta \kappa}{\gamma} \; \; \; = \; \; \; - \frac{1}{2}
\frac{\delta I}{(I_c)^3} \; - \; \frac{\lambda^c}{\gamma}
\frac{M_c}{I_c} \left[ \; \frac{\delta M}{M_c}  -
\frac{\delta I}{I_c} \right] \; \; \; = \; \; \;
\left[ \frac{- \pi^2}{256} + \frac{4}{3} \frac{\lambda^c}{\gamma} \right]
\frac{\delta I}{I_c} \; \; - \; \;
\frac{\lambda^c}{\gamma} \frac{\varepsilon}{2} \; ,
\label{Cvar1}
\end{equation}
where in the last step we made use of the relation between the
variations $\delta I$ and $\delta M$ and inserted the explicit
results for $I_c = I(1)$ and $M_c = M(1)$.
Using the result for $\delta I/I_c$ to
lowest order we obtain (see~\cite{GaPaSe97} for further details)
\begin{equation}
\delta \kappa  \; \; = \; \; - \; \varepsilon
\ln (\varepsilon) \; \sigma \; ,
\label{Cvar2}
\end{equation}
where we introduced the abbreviation
\begin{equation}
\sigma \; \equiv \; \frac{1}{8} \left[- \frac{\gamma \pi^2}{256} +
\frac{4 \lambda^c}{3} \right] \; .
\label{Csigma}
\end{equation}
Inverting equation (\ref{Cvar1}) (again taking into account only the
leading order) gives
\begin{equation}
\varepsilon \; = \; - \; \sigma^{-1} \delta \kappa \;
\left[ \ln \left( \sigma^{-1} \delta \kappa
\right) \right]^{-1}\; .
\label{Cinv}
\end{equation}
This equation is the relation between the variation $\delta
\kappa$ and $\varepsilon$ which is implied by the necessary
condition (\ref{Cnecc}). In the final step we expand the derivative of the
free energy (\ref{Cdfmu}) at
$\mu = 1$ and use the result (\ref{Cinv}) to obtain the variation of the
derivative in terms of $\delta \kappa $. Expanding
(\ref{Cdfmu}) gives
\begin{equation}
\frac{d  f}{d \kappa } \; \; \; = \; \; \; -2 \frac{M_c}{I_c}
\left[ 1 + \frac{\delta M}{Mc} - \frac{\delta I}{I_c} \right]
\; \; \; = \; \; \;
-\frac{2}{3} \; - \; \frac{1}{3} \varepsilon \ln(\varepsilon)
\; - \; \varepsilon \; .
\label{Cfin1}
\end{equation}
Using (\ref{Cinv}) we obtain
\begin{equation}
\frac{d  f}{d \kappa} \; \; \; = \; \; \;
-\frac{2}{3} \; \; + \; \; \frac{1}{3} \sigma^{-1} \delta
\kappa \; \; + \; \;
\sigma^{-1} \delta \kappa  \; \left[ \ln \left( 
\sigma^{-1} \delta \kappa \right) \right]^{-1} \; .
\label{Cfin2}
\end{equation}
We remark, that the case $\mu < 1$ with expansion $\mu = 1 - \varepsilon,
\; \varepsilon > 0$ changes only the sign of the argument of the
logarithm. Differentiating the last result with respect to $\delta \kappa$
establishes the singular behaviour of the third derivative of the free
energy with respect to $\kappa$. Thus we find a third order phase transition
for $\mu = 1$. The critical line is a straight line given by
(\ref{Ccritlin}).

It is important to notice, that at (see Equation (\ref{Cvar1}))
\begin{equation}
\frac{\lambda^c}{\gamma} \; =  \;\frac{3 \pi^2}{1024} \; ,
\label{Clamterm}
\end{equation}
the leading term in the
expression for $\delta \kappa$ vanishes. Equation (\ref{Cvar1})
is reduced to the simpler relation
\begin{equation}
\frac{\delta \kappa }{\gamma} \; = \; - \varepsilon \;
\frac{\lambda^c}{2 \gamma} \; .
\label{Cvar3}
\end{equation}
At this point the expansion of $df/d\kappa$ gives
\begin{equation}
\frac{d f}{d \kappa } \; = \; -\frac{2}{3} \; - \;
\frac{1}{3} \varepsilon \ln(\varepsilon) \; - \; \varepsilon \; = \;
-\frac{2}{3} \; + \; \frac{2}{3 \lambda^c}  \delta \kappa
\ln\left(- \frac{2}{\lambda^c} \delta \kappa \right) \; + \;
\frac{2}{\lambda^2} \delta \kappa  \; .
\label{Cfinscnd}
\end{equation}
Again the case $\mu < 1$ differs only by the sign of the argument of the logarithm. Differentiation with respect to $\delta \kappa$
shows, that the phase transition has turned to second order at that point.
Using (\ref{Ccritlin}) one can compute also the $\kappa/\gamma$
coordinate of the second order point giving $\kappa/\gamma = \pi^2/1024,
\lambda/\gamma = 3 \pi^2/1024$. In fact the more global analysis in the
next section will show, that the third order line terminates at
the second order point $\kappa/\gamma = \pi^2/1024,
\lambda/\gamma = 3 \pi^2/1024$, and continues as a first order line.

\subsection{Regime of the first order phase transition}

\begin{figure}[htbp]
\hspace*{18mm}
\psfig{figure=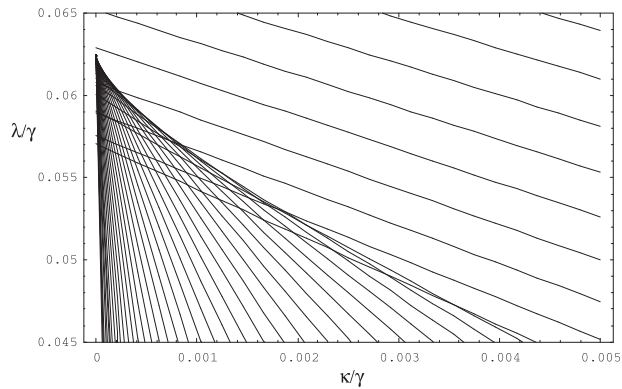,height=2in}
\caption{Plot of the lines (3.18) for $\mu$ ranging from 
$0.4$ (upper right corner) to $75$ (line at the extreme left).  
The region of overlapping lines corresponds to
a region of first order phase transition. \label{linesfig}}
\end{figure}

As pointed out at the end of the last section at the point $\kappa/\gamma
= \pi^2 / 1024$, $\lambda/\gamma = 3 \pi^2 / 1024$, the third order
transition along the $\mu=1$ line (\ref{Ccritlin}) changes to second order.
This unusual behaviour requires further investigation which we will
carry out in this section. To begin, a graphical analysis
of the phase diagram is most useful and in Figure
\ref{linesfig} we plot a number
of representatives of the family of lines (\ref{Cnecc}) for a range of
values of $\mu$.

It is clear that in most of the $\kappa, \lambda$-plane
points are in a one-to-one correspondence with values of the parameter $\mu$.
This correspondence breaks down though in a small
region near the $\lambda / \gamma$ axis between $\lambda / \gamma =0.05$
and $\lambda / \gamma =0.0625$.
Due to the behaviour of the slope and intercept in the linear equation
(\ref{Ccritlin}) lines begin to overlap for increasing $\mu$ starting at
$\mu \sim 1$ and continuing as $\mu \rightarrow \infty$.
In this overlap region the phase diagram is folded at the vertex
$\kappa/\gamma= \pi^2 / 1024$, $\lambda/\gamma = 3 \pi^2 / 1024$, and
each point falls on three  different
lines of constant $\mu$. Consequently the system simultaneously
admits three configurations with different free energies in
this region of the phase space. This circumstance
allows for a first order phase transition to develop along a line where the
free energies of the different phases are equal.
\begin{figure}[htbp]
\hspace*{17mm}
\psfig{figure=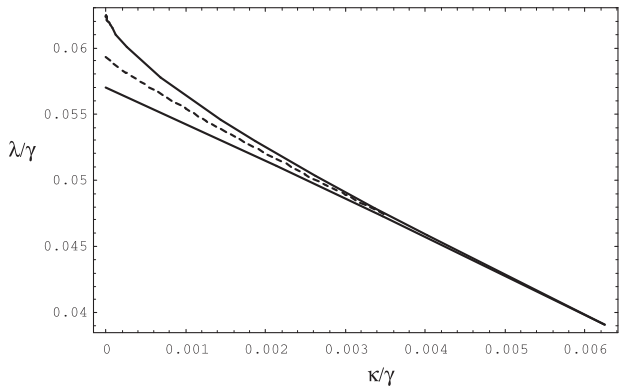,height=2in}
\caption{Plot of the boundary of the multiple phase region.
The boundary (solid curve) is given by a caustic of lines in the one-parameter
family (3.18). The dotted curve shows the numerically determined 
first order line.
\label{causticfig}}
\end{figure}

The edges of the triangular first order region in Figure \ref{linesfig}
is given by a caustic of lines from the
one parameter family (\ref{Cnecc}).  The boundary is defined by the curve where
the family of curves is stationary with respect to $\mu$. This condition can
be used with (\ref{Cnecc}) to give a definition of the boundary caustic
shown in Figure 2.
The stationary condition can be solved with the parametric result
\beq
\frac{\kappa}{\gamma} \; \; = \; \;
\frac{1}{4 I^2(\mu) } \; \frac{ I^{\prime}(\mu) M(\mu) +
M^{\prime}(\mu) I(\mu) }{M^{\prime}(\mu) I(\mu) -I^{\prime}(\mu) M(\mu) }
\label{caustic} \; .
\eeq
As can be seen, the curve given by (\ref{caustic}) intersects the
$\lambda / \gamma$
axis at two points: $0.057024~ (\mu = 0.95324)$ and $1/16~(\mu= \infty)$
and reaches a singular maximum in the $\kappa / \gamma$ direction
for $\mu = 1$ at the point $\kappa / \gamma= \pi^2 /1024$.  The end of
this region of first order transitions agrees with the position of the second
order transition point which was determined by the analysis of critical behaviour in the previous section.

Once one has determined the region where different phases can co-exist the
next issue to address is that of the position of the line of first order
phase transitions where different phases have the same free energy.
\begin{figure}[htbp]
\hspace*{21mm}
\psfig{figure=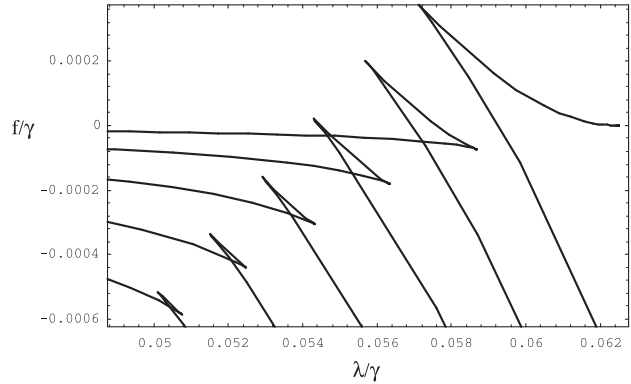,height=2in}
\caption{Free energy $f/\gamma$ as a function 
of $\lambda/\gamma$ in the region of first order phase transitions.
Each curve is plotted for fixed $\kappa / \gamma$ which from right to left
is $\kappa / \gamma  =
0,~0.0005,~0.001,~0.0015,~0.002,~0.0025$ \label{freeE} }
\end{figure}

The first order line can be determined for given 
$\kappa / \gamma$ by the simultaneous
solution for the parameters $\mu_1$ and $\mu_2$ of the pair of equations
\beq
\frac{I(\mu_2) M(\mu_2) - I(\mu_1) M(\mu_1)}{ 4 \; I(\mu_1) I(\mu_2)} \; =
\; \frac{ \kappa}{\gamma} \Big[
M(\mu_2) I(\mu_1) - I(\mu_2) M(\mu_1) \Big] \; ,
\label{firsto1}
\eeq
and
\beq
\frac{1}{6} \left( \frac{\mu_1}{I^2(\mu_1)} -
\frac{ \mu_2}{I^2(\mu_2)} \right) -\frac{1}{12}
\left( \frac{M(\mu_1)}{I^3(\mu_1)} - \frac{ M(\mu_2)}{I^3(\mu_2)} \right)
\; = \;
\frac{\kappa}{\gamma} \frac{I(\mu_2) M(\mu_1) - M(\mu_2) I(\mu_1)}
{I(\mu_1) I(\mu_2)} \; .
\label{firsto2}
\eeq

Unfortunately, these equations are analytically intractable.  Again we
turn to a graphical analysis to gain further insight.  In Figure \ref{freeE}
we plot the free energy of the system as a function of $\lambda / \gamma$
for different values of fixed $\kappa / \gamma$.  From here it is easy to
see a number of features of the region of first order transitions.
Increasing $\mu$
traverses these curves in a clock-wise rotation so that the 
free energy increases
for small values of $\mu$, intersecting the nearly horizontal
large $\mu$ free energy.  This intersection point is a graphical
demonstration of the first order transition which occurs here
as the model jumps from weak $(\mu<1)$ to strong coupling $(\mu$ large$)$.
Each phase continues to exist after the transition point and may be
reached by an adiabatic process until ending in cusps which mark the
boundaries of the first order region in the $\lambda / \gamma$ axis.
It is interesting to note that there is an energetically unfeasible
intermediate ``medium coupling'' phase which connects the weak and strong
phases. Hence, for fixed $\kappa / \gamma$ there exist three distinct
configurations of the system for given $\lambda / \gamma$ in the
region of first order transitions.

\subsection{Phase diagram}
We end this section with a short discussion of the phase diagram 
analyzed in the previous sections. Figure 4 shows a schematic picture
of the phase diagram.
\begin{figure}[htbp]
\hspace*{34mm}
\psfig{figure=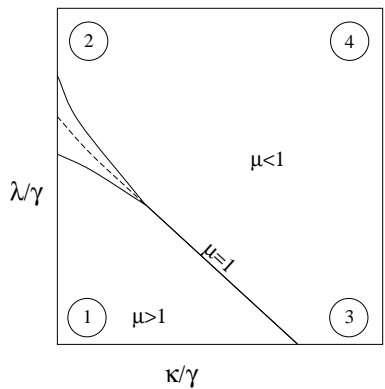,height=2in}
\caption{Schematic picture of the phase diagram. 
The dashed curve marks the first order 
part of the critical line. The solid curves above and below it are the 
boundaries of the area with two possible phases. They join at a point
which shows second order behaviour. For larger $\kappa/\gamma$
we find a third order line ($\mu = 1$) marked by a solid line. The numbers
label four extremal corners of the phase diagram.
We also indicate the range of the auxiliary parameter $\mu$.
\label{phasediag}}
\end{figure}
The dashed line gives the numerically determined first order
line. The full curves above and below show the boundaries of the region where
different phases can coexist. With increasing $\kappa/\gamma$, this region 
narrows, and ends in a point which is second order as discussed above. The
curve separating the two phases continues as the third order line  
(\ref{Ccritlin}). 

We also show the range of the auxiliary parameter $\mu$. In the hot and 
dense (deconfining) phase it assumes values $\mu < 1$, while in the 
cold and dilute (confining) phase it is restricted to $\mu > 1$. 
On the third order part of the critical line we have $\mu = 1$. 

It is interesting to discuss the extremal regions of the phase diagram: 
In the case of only adjoint charges
(all $k_n = 0$) the action is invariant under transformations in the center
of the gauge group, $g \rightarrow z g$ (compare (\ref{zsym})). When this symmetry is faithfully represented we expect
\begin{equation}
c_n(x) \; = \; \lim_{N \rightarrow \infty} \frac{1}{N} \langle {\rm
Tr}~g^n(x) \rangle \; = \; 0 \; \; \; \; \; \; \forall n \neq 0 \; ,
\label {consol}
\end{equation}
since
\[
{\rm
Tr}~g^n(x)~\rightarrow z^n~{\rm Tr}~g^n(x) \; .
\] From (\ref{coll1}), (\ref{coll2}) it is obvious that (\ref{consol})
which corresponds to constant eigenvalue density $\rho \equiv 1/2\pi$ 
is a solution. The stability
analysis performed in \cite{stz} shows that this solution ("confined
phase") is stable for $\lambda/\gamma \in [0, 1/16]$. Since the upper
boundary of the multiple phase region
is at $\lambda/\gamma = 1/16$ we
confirm this result.

The center symmetry of the action is explicitly broken when there are
fundamental charges (some $\kappa_n \neq 0$). In particular this is the 
case when we consider the model with only fundamental charges (and their
conjugate). Thus we cannot expect to find a confining solution with vanishing 
$c_n$ which would correspond to vanishing (energy) densities  of 
colour electric string $(\rho_{Flux})$, fundamental representation
charges 
$(\rho_{F,\bar{F}})$ and adjoint representation charges 
$(\rho_{Ad})$
(please see \cite{GaPaSe97} for details). 
For this case ($\lambda = 0$) we have
established the existence of a third order 
phase transition at $\kappa/\gamma =  \pi^2/512$ 
(set $\mu=1, \lambda = 0$ in (\ref{Cnecc})). 

Finally we discuss extremal corners of the phase diagram, which
are labeled 1,..,4 in Figure \ref{phasediag}. 
The four extremal cases can easily be understood by the
magnitude of the energy densities $\rho_{Flux}$, $\rho_{Ad}$ and
$\rho_{F, \bar{F}}$.
Point 1 is in the extremal corner of the low density phase. 
All three densities are rather small. Points 2 and 3 are both in the
high density phase, in areas which are dominated by adjoint charges
(Point 2) and fundamental charges (Point 3). It is nice to see,
how the energy density of the sources is dominated by the contributions 
of the adjoint charges and fundamental charges respectively. Finally
Point 4 is in a region where both the density of the adjoint charges
as well as the fundamental charges is high and of the same magnitude.

\section{Summary}
In this paper we analyzed the thermodynamic properties of a model of static
sources on a line
interacting through non-Abelian forces. It was shown that the partition
function takes the form of the partition function of a gauged 
principal chiral model. Using the eigenvalue density as collective field
variable the Hamiltonian for the eigenvalue density in the large $N$-limit 
was computed. We gave a
static solution of the corresponding Hamilton equations. For the special case
of only two types of charges, the static solution 
was parametrized using the parameter $\mu$ proportional to the Fermi energy. 
In particular the case of two types of 
charges transforming under the adjoint, and charges 
transforming under the fundamental representation of the gauge group was 
considered. 
Expanding the parametrized solutions at $\mu = 1$ we established the 
existence of a straight line in the phase diagram where the free 
energy exhibits a third order phase transition. We proved that the third
order behaviour terminates at a second order point. The critical line
then continues as a first order line, which was determined numerically.

The whole phase diagram was interpreted by analyzing the contributions
of charges and flux to the energy density. We found that 
for $\mu >1$ the system is characterized by low energy densities, while
for $\mu < 1$ the densities are high. 

This 2-dimensional model could be generalized in several directions. 
It would be interesting to analyze non-static solutions of the Hamilton
equations and different boundary conditions which might be used to 
include a $\theta$-term. Loop expansion of the fermion determinant of 
QCD$_2$ with large quark masses could be used to relate the fugacities 
of the non-Abelian gas analyzed in this article to the mass parameters
of QCD$_2$.

\vspace*{-2pt}

\subsection*{Acknowledgment}
This work was supported in part by the Natural Sciences and Engineering
Research Council of Canada. L.~P.~is supported 
in part by a University of British Columbia Graduate Fellowship.

\eject

\end{document}